\documentclass[twocolumn,11pt]{article}

\usepackage[T2A]{fontenc}
\usepackage[utf8]{inputenc}
\usepackage[english]{babel}
\usepackage{amsmath}
\usepackage{amssymb}
\usepackage{textcase}
\usepackage{bm}
\usepackage{multicol}
\usepackage{tabularx}
\usepackage{euscript}
\usepackage{calc}
\usepackage{cite}
\usepackage{indentfirst}
\usepackage{mathtext}
\usepackage{titlesec}
\usepackage{graphicx}
\usepackage[a4paper, left=1.75cm, right=1.5cm, top=1.9cm, bottom=2.6cm]{geometry}

\titleformat*{\section}{\centering\large\uppercase}

\def\beq{\begin{equation}}
\def\eeq{\end{equation}}

\begin{document}

\twocolumn[
\begin{@twocolumnfalse}
\vspace{3em}
\centering
\begin{minipage}{0.84\textwidth}

\begin{center}
{\Large{\bf 
 Insulator-bad metal transition in RNiO$_3$ nickelates beyond  Hubbard model and density functional theory
}}

\vspace{1em}
\textbf{\large 
A.~S. Moskvin{*\textsuperscript{,}**}
}
	
{\small
{*}\textit{Ural Federal University,
	620083 Ekateringburg, Russia}
	
{**}\textit{M.N. Mikheev lnstitute of Metal Physics of Ural Branch of Russian Academy of Sciences,
	620108 Ekateringburg, Russia}

\vspace{1em}
{*}\textit{e-mail: alexander.moskvin@urfu.ru}
}

\end{center}

{\small
The insulator-bad metal transition observed in the Jahn-Teller (JT) magnets orthonickelates RNiO$_3$ (R = rare earth or yttrium Y) is considered to be a canonical example of the Mott transition, traditionally described in the framework of the Hubbard $U$-$t$-model and the density functional theory. However, actually the real insulating phase of nickelates is the result of charge disproportionation (CD) with the formation of a system of spin-triplet (S=1) electron [NiO$_6$]$^{10-}$ and spinless (S=0) hole [NiO$_6$]$^{8-}$ centers, equivalent to a system of effective spin-triplet composite bosons moving in a nonmagnetic lattice. Taking account of only charge degree of freedom we develop a novel minimal $U$-$V$-$t_b$-model for nickelates making use of  the charge triplet model with the pseudospin formalism and  effective field approximation. We show the existence of two types of CD-phases,  high-temperature classical CO-phase with the G-type charge ordering of electron and hole centers, and low-temperature quantum CDq-phase with charge and spin density transfer between electron and hole centers, uncertain valence and spin value for NiO$_6$ centers. Model $T$-R phase diagram reproduces main features of the phase diagram found for RNiO$_3$.

\vspace{1em}
}
\vspace{2em}

\end{minipage}

\end{@twocolumnfalse}
]

{\bf I.\,Introduction.} Nickelates RNiO$_3$ (R = rare earth or yttrium Y) demonstrate extremely unusual electrical and magnetic properties, firstly it is a weakly first-order metal-insulator transition (MIT) observed in orthorhombic RNiO$_3$ (R = Lu,..., Pr)  upon cooling below $T_{\mbox{{\footnotesize MIT}}}$ in the range from 130\,K for Pr to $\sim$\,550-600\,K for heavy rare earths\,\cite{Medarde_1997,Catalano_2018,Gavrilyuk}.
For more than three decades, the origin of MIT in nickelates has challenged the condensed matter research community.

Nickelates RNiO$_3$ belong to a broad class of Jahn-Teller (JT) magnets\,\cite{Khomskii_2014,Moskvin_2013,Moskvin_2023,Moskvin_2025}, compounds based on Jahn-Teller $3d$- or  $4d$-ions  with $t_{2g}^{n_1}e_g^{n_2}$-type configurations in highly symmetric octahedral, cubic or tetrahedral environments and with an orbital $E_g$-doublet in the ground state. 

All the JT magnets are strongly correlated systems in the sense that they cannot be adequately described within the framework of density functional theory (DFT), or more precisely DFA (density functional approximation), as it is an ill-defined  questionable "starting"\, approximation for describing local and non-local correlations, which practically excludes their adequate consideration (see, e.g.,\,\cite{Moskvin_OS_2016}). 
Even the use of the most advanced \textit{ab initio} DFT-based methods as LDA+DMFT and GGA+DMFT does not provide an adequate self-consistent description of the lattice, orbital and spin ordering for such a prototypical JT insulator as KCuF$_3$\,\cite{Leonov_2008,Pavarini_2008}.

Much more problematic is the application of these methods to the description of JT magnets unstable to charge transfer, such as nickelates RNiO$_3$, with their characteristic metal-insulator transitions. 
The traditional approach to describing MIT with a sharp change in the magnitude and temperature dependence of resistance assumes the implementation of a spontaneous "order-to-order"\, phase transition from the high-temperature coherent Fermi liquid phase that admits a more or less adequate DFT description  to a low-temperature insulating phase with a charge ordering.
A typical theoretical model of MIT is the Hubbard $U$-$t$ model\,\cite{Imada}, which takes into account the single-particle kinetic energy determined by the effective transfer integrals $t$ and local correlations determined by the effective parameter $U$. As a rule, the transport integrals are found within the DFT-based tight-binding approximation and $U$ is treated as a fitting parameter for $d$-electrons.
However, for the vast majority of JT magnets, the high-temperature metal-like (bad metal) phase is fundamentally different from the coherent Fermi liquid and is more reminiscent of "... a nondegenerate gas of small polarons interacting through very weak Coulomb forces, but which would condense in some way, at a low enough temperature" (see review paper by Mott\,\cite{Mott_1}).  Obviously, such a phase cannot be adequately described within the DFT-based approach.


Despite the numerous experimental findings of the non-Fermi-liquid  behaviour of the high-temperature bad metallic phase in nickelates, the MIT in RNiO$_3$ has been considered as a canonical example of a "band-width-driven"\, transition\,\cite{Imada} usually  described within the effective two-band Hubbard model (see, e.g., Refs.\,\cite{Subedi_2015,Lu}).  
However, the MIT in nickelates has been considered by different researchers with varying conclusions regarding the driving force behind the transition, from  different pure electronic DFT-based mechanisms\,\cite{Subedi_2015,Lu,Park_2012,Varignon} to  an unexpected result of only the coupling between lattice modes \cite{lattice}. Variety of the most recent DFT-based approaches  converged to a common picture  with a key role played by the electron-lattice interaction, bond and charge disproportionation\,\cite{Catalano_2018,Peil_2019,Millis_2022}. However, the question of the reliability of these results remains open\,\cite{Millis_2022}. Many unsolved and unsoluble problems give rise
to serious doubts in quantitative and even qualitative predictions made within the DFT-based approaches.

Despite the optimistic statements of many theorists (see, e.g., \,\cite{Varignon}) and even experimentalists\,\cite{Catalano_2018}, standard DFT-based methods (LDA, LDA+U, LDA+DMFT,...)  do not and cannot describe all  the complex of basic properties of nickelates, although it cannot be denied that they can give, at first glance, a quite plausible semi-quantitative description of some individual characteristics of the electronic structure of these strongly correlated systems. The adequate description of the MIT and other properties of nickelates requires going beyond Hubbard model and DFT-based methods.

Today, few researchers doubt that the low-temperature insulating phase of nickelates is the result of charge disproportionation (CD), but the question of the structure of the CD phase, key interactions and the effective Hamiltonian remains open. 
We believe that the low-energy physics of the charge-disproportionated nickelates is determined by a system of charge triplets, including a  bare JT-center [NiO$_6$]$^{8-}$, spin-triplet (S=1) electron [NiO$_6$]$^{10-}$ and spinless (S=0) hole [NiO$_6$]$^{8-}$ centers\,\cite{Moskvin_2013,Moskvin_2023}.  
 We argue that local ($U$), nonlocal ($V$) correlations, and two-particle transfer integral ($t_b$) govern     the charge degree of freedom in nickelates. 
Making use of  the charge triplet model with the pseudospin formalism and  effective field approximation we develop a novel minimal $U$-$V$-$t_b$-model for nickelates that provides an adequate description of the insulator-to-bad-metal phase transition as melting of the CD phase and the famous $T_{\mbox{{\footnotesize MIT}}}$-R phase diagram found for RNiO$_3$\,\cite{Catalano_2018,Gavrilyuk}. 
We show the existence of two types of CD-phases, high-temperature classical CO-phase with the G-type charge ordering of electron and hole centers, and low-temperature quantum CDq-phase with charge and spin density transfer between electron and hole centers, "uncertain valence"\, [NiO$_6$]$^{9\pm\delta}$, or Ni$^{3\pm\delta}$ (0$\leq\delta\leq$1) and spin value 0$\leq\langle S\rangle\leq$1. 
 
 The paper is organized as follows. First, in Sec.\,II we describe the disproportionation phenomenon in nickelates. Then, in Secs.\,III--V  we introduce  the charge triplet model, pseudospin formalism and effective Hamiltonian with a focus on the two-particle, or bosonic transfer. Finally, in Secs.\,VI and VII we present a short information on the effective field model and   results of calculations of $T$-$V$ and $T$-$t_b$ phase diagrams within the $U$-$V$-model (atomic limit) and extended $U$-$V$-$t_b$-model. In Sec.\,VIII, we summarize our main results. In Supplemental Materials we present a more detailed textbook description of the effective field theory in relation to our model and a brief note on magnetic structure.

{\bf II.\,Anti-JT disproportionation in nickelates.} All JT configurations of $d$-ions include one $e_g$-electron or one $e_g$-hole over stable, fully or half-filled, shells. In this sense, they are similar to the configurations of the numerous family of ions with one $ns$-electron over filled shells, such as the $6s$-electron in Hg$^+$, Tl$^{2+}$, Pb$^{3+}$, Bi$^{4+}$. These ionic configurations are unstable with respect to the disproportionation reaction, or even nonexistent (missing oxidation states\,\cite{Katayama}).Thus, in BaBiO$_3$, instead of the nominal valence 4+, bismuth prefers the stable valence states Bi$^{3+}$ and Bi$^{5+}$ with fully filled shells. However, unlike ions with $ns$-electrons, for JT ions we deal with orbital degeneracy for $e_g$-electrons/holes, hence the possibility of competition between the Jahn-Teller effect leading to orbital ordering\,\cite{Khomskii_2014}, and the anti-JT disproportionation effect, leading to the formation of a system of $S$-type electron and hole centers with an orbitally nondegenerate ground state\,\cite{Moskvin_2013,Moskvin_2023}, equivalent to a system of effective composite spin-singlet or spin-triplet bosons in a nonmagnetic (“single-band”\, JT-magnets) or magnetic (“two-band”\, JT-magnets) lattice\,\cite{Moskvin_2023}.
The stability of the JT phase is determined both by the large positive value of the local correlation energy $U$, the JT stabilization energy\,\cite{Khomskii_2014} and by the strong vibronic reduction of single-particle transport integrals.

The ESR (electron spin resonance)\,\cite{LaAlO3_JT_ESR} and optical spectra measurements\,\cite{LaAlO3_JT} revealed clear signatures of JT effect associated with a low-spin (LS) ground state of the impure Ni$^{3+}$ ions in the perovskite LaAlO$_3$.  
However, nominally Ni$^{3+}$ ions show no significant traces of JT-distortion and orbital order (OO) in nickelates  RNiO3 with the close to LaAlO$_3$ perovskite structure\,\cite{no-OO}, both below and above the MIT transition temperature, that is an indirect signature to charge disproporionation in nickelates. 

\begin{table*}[t]
		\label{tab1}
		\caption{{\bf Tab.\,1.} Pseudospin, spin and orbital structure of three NiO$_6$ charge centers in RNiO$_3$ orthonickelates}
		\centering
		\begin{tabular}{|c|c|c|c|c|c|}
\hline
$d$-center & Ion & Cluster & Pseudospin projection & Spin & Orbital state \\
\hline
Electron, $d^8$ & Ni$^{2+}$ & [NiO$_6]^{10-}$ & M\,=\,--1 & $S$\,=\,1 & $t_{2g}^6e_g^2;{}^3A_{2g}$ \\ \hline
Jahn-Teller, $d^7$ & Ni$^{3+}$ & [NiO$_6]^{9-}$ & M\,=\,0 & $S$\,=\,1/2 & $t_{2g}^6e_g^1;{}^2E_{g}$ \\ \hline
Hole, $d^6$ & Ni$^{4+}$ & [NiO$_6]^{8-}$ & M\,=\,+1 & $S$\,=\,0 & $t_{2g}^6;{}^1A_{1g}$ \\ \hline
       \end{tabular}
	\end{table*}

Anti-Jahn-Teller symmetric $d$-$d$-disproportionation according to the scheme
$$
\mbox{Ni}^{3+} + \mbox{Ni}^{3+} \rightarrow \mbox{Ni}^{2+} + \mbox{Ni}^{4+}  \Leftrightarrow \mbox{Ni}^{4+} + \mbox{Ni}^{2+}  
$$
\beq
\left(d^7 + d^7 \rightarrow d^{8} + d^{6} \Leftrightarrow d^{6} + d^{8}\right)
\label{dis1}
\eeq
involving the forming  a system of electron $d^{9}$ and hole $d^{6}$ centers differing by an electron/hole pair, is an alternative and competing mechanism for the removal of orbital degeneracy for Ni$^{3+}$($t_{2g}^6e_g$) ions\,\cite{Moskvin_2013,Moskvin_2023,Moskvin_1998,Mazin,Moskvin-LTP,Moskvin-09,Moskvin_2011} and the suppression of the JT effect in nickelates. 
Obviously, in systems with strong $d$-$p$-hybridization (cation-anion covalence), the disproportionation reaction instead of a naive ionic picture (\ref{dis1}) should be written in a “cluster”\ language as follows
$$
[\mbox{NiO}_6]^{9-}+[\mbox{NiO}_6]^{9-}\,\rightarrow\,
$$
\beq
[\mbox{NiO}_6]^{10-}+[\mbox{NiO}_6]^{8-} \Leftrightarrow [\mbox{NiO}_6]^{8-}+[\mbox{NiO}_6]^{10-}\, .
\label{dis2}
\eeq
The cluster formalism is actually a generalization of ligand field theory, in which single-particle $t_{2g}$ and $e_g$ states are described as molecular orbitals, or linear combinations of the atomic $3d$ orbitals of the cation and the $2p$ orbitals of the ligands. This approach immediatelly takes into account the local point symmetry of the cation and generalizes the well-known and still popular atomic-molecular approach (see, e.g, \,\cite{Johnston}), in which the clusters  are composite objects described by the combination of atomic states
$$
\mbox{Ni}^{3+}\rightarrow[\mbox{NiO}_6]^{9-}:  3d^7; 3d^8\underline{L}; 3d^9\underline{L}^2\, ,
$$
$$
\mbox{Ni}^{2+}\rightarrow[\mbox{NiO}_6]^{10-}:  3d^8; 3d^9\underline{L}; 3d^{10}\underline{L}^2\, ,
$$
$$
\mbox{Ni}^{4+}\rightarrow[\mbox{NiO}_6]^{8-}:  3d^6; 3d^7\underline{L}; 3d^8\underline{L}^2\, ,
$$
where $\underline{L}$ denotes a ligand hole. 
 The cluster model assumes replacement of the cation-anion system in nickelates by a lattice of  NiO$_6$ clusters (Ni$^{2+,3+,4+}$-centers), whose Wannier-type wave functions  effectively account for the $p$-$d$-covalence and the "negative charge transfer energy"\, effects\,\cite{Johnston}.  
Let us pay special attention to the right part of the expressions (\ref{dis1})-(\ref{dis2}), indicating the possibility of realization of both classical ("chemical") disproportionation with the forming  centers with a certain charge $n\pm 1$, and quantum ("physical") disproportionation with the forming quantum superpositions of $n\pm 1$-centers with uncertain valence. The electron ($n+1$) and hole ($n-1$) centers  are distinguished by a pair of electrons/holes, which represent an effective composite boson. 

Modern concepts of disproportionation in nickelates are usually limited to the idea of the classical system of electron and hole centers such as Ni$^{2+}$ and Ni$^{4+}$ or Ni$^{3\pm\delta}$   as a result of negative values of the parameter of local correlations $U$ (negative-$U$ model) or "negative charge transfer energy $\Delta$"\, (see, e.g., \,\cite{Mazin, Johnston,Medarde_2009}) without any analysis of the fundamental role of nonlocal correlations and two-particle (bosonic) transfer in the formation of the electron and magnetic structure.


{\bf III. Charge triplet model: pseudospin formalism.} Following the remarkable idea of Rice and Sneddon\,\cite{Rice_1981} developed by us for 2D cuprates and other JT magnets\,\cite{Moskvin_2013,Moskvin_2023,Moskvin_2011,ASM_CM_2021,ASM_JMMM_2022}, we propose a generalized model of effective charge triplets to describe the electron structure and phase diagrams of RNiO$_3$, which implies consideration of some highly symmetric “parent”\ configuration with ideal octahedra NiO$_6$, whose low-energy state is formed by a charge triplet [NiO$_6$]$^{10-,9-,8-}$ (nominally Ni$^{2+,3+,4+}$) with different spin and orbital ground states. We associate the three charge states of a NiO$_6$ cluster with three pseudospin $\Sigma$\,=\,1 projections  and use the well-known spin algebra and other methods, well established for spin magnets, to describe the charge degrees of freedom for nickelates in the “coordinate”\, representation instead of the traditional ${\bf k}$-representation for the DFT-based models. The pseudospin, spin and orbital structure of the three charge NiO$_6$-centers   in nickelates  are presented in Table\,1. In the simplest approximation, below we neglect both the possible difference of the $d$-$p$-structure of the single-particle $t_{2g}$- and $e_g$-states for different components of the charge triplet and the contribution of the inactive $S$\,=\,0 fully filled $t_{2g}^6$-shell to various spin and orbital interactions.

Formally, the local pseudospin $\Sigma$\,=\,1 implies eight (three “dipole”\, and five “quadrupole”\,) independent operators and corresponding parameters of the local charge order. In the irreducible components these are 
$$
\Sigma_0=\Sigma_z; \Sigma_{\pm}=\mp \frac{1}{\sqrt{2}}(\Sigma_x\pm i\Sigma_y); \Sigma_z^2; \Sigma_{\pm}^2; T_{\pm}=\frac{1}{2}\{\Sigma_z,\Sigma_{\pm}\} \, .
$$
The value $n_{e_g}$=\,1\,-\,$\langle {\hat \Sigma}_z\rangle$ is the mean on-site number of $e_g$-electrons, $\Delta n$\,=\,$\langle {\hat \Sigma}_z\rangle$ determines the deviation from half-filling. 
  The operators $P_0=(1-\Sigma_z^2)$; $P_{\pm}=\frac{1}{2}\Sigma_z^2(1\pm\Sigma_z)$ are actually projection operators on charge states with pseudospin projection M = 0, $\pm$1, respectively, and the mean $\langle P_0\rangle$, $\langle P_{\pm}\rangle$ are actually local densities for the corresponding charge states.   
 The operators $\Sigma_{\pm}$ and $T_{\pm}$ change the pseudospin projection to $\pm$1. The operators $\Sigma_{\pm}^2$ change the pseudospin projection to $\pm$2, so these can be viewed as the creation/annihilation operators for the effective composite boson as superconducting carriers\,\cite{RMP}. The corresponding local averages $\langle \Sigma_{\pm}\rangle$, $\langle T_{\pm}\rangle$, $\langle \Sigma_{\pm}^2\rangle$ will describe different variants of "off-diagonal"\, charge order, in particular, coherent metallic and superconducting states. 
  
  Given the spin and orbital states for the charge components, we must extend the on-site Hilbert space to a "pseudospin-orbital-spin octet"\, 
$$
|\Sigma M;\Gamma\mu;Sm\rangle = |1M;\Gamma\mu;Sm\rangle \, ,
$$
($\Gamma = A_{1g}, A_{2g}, E_g$ is irreducible representation of local point group) including the Jahn-Teller spin-orbit quartet $|10;E_g\mu;\frac{1}{2}\nu\rangle$ with M\,=\,0
and spin-charge quartet with M\,=\,$\pm$1, including a singlet $|1+1;A_{1g}0;00\rangle$ and a triplet 
 $|1-1;A_{2g}0;1m\rangle$, where $\mu =\,0,\,2$, $\nu =\,\pm\,\frac{1}{2}$, $m\,=\,0,
\,\pm 1$ ($|E_g0\rangle \propto d_{z^2}; |E_g2\rangle \propto d_{x^2-y^2}$), and consider a low-energy physics for nickelates to be formed by  a system of such octets. This approach allows us to take into account in the most general form the effects of competition of different degrees of freedom.

In order to develop a minimal model of the low-energy electronic structure for nickelates, below  we will focus on the consideration of only the charge degree of freedom, by design neglecting the electron-lattice and superexchange interactions, as well as the effects of the vibronic-reduced single-particle transport. In this approximation, the on-site structure of the NiO$_6$-octets will be reduced to a fourfold degenerate Jahn-Teller level with $M$\,=\,0, a singlet with $M$\,=\,+1, and a spin triplet with $M$\,=\,-1. Additionally, we will limit ourselves to considering the cubic perovskite structure of Ni-centers and taking into account the $nn$-interaction of nearest neighbors. Such a minimal model describes not the Jahn-Teller, but rather the “anti-Jahn-Teller” physics of nickelates, as JT magnets unstable with respect to charge transfer with the formation of CD phases.

{\bf IV. Effective low-energy Hamiltonian}. 
The effective Hamiltonian of NiO$_6$ centers to describe the charge degree of freedom will be reduced to only three terms, the energy of local and nonlocal charge correlations and two-particle transport. 
The effective Hamiltonian of the isolated NiO$_6$ center  includes only local correlations 	
\begin{equation}
	{\hat H}_{loc} = \frac{U}{2}\sum_{i} {\hat \Sigma}_{iz}^2
\label{loc}	
\end{equation}
to be an analog of the single-ion axial spin anisotropy that describes the effects of a "bare"\, pseudospin splitting.  Positive values of the local correlation parameter $U$\,>\,0 stabilize the Jahn-Teller spin-orbit quartet $|10;E_g\mu;\frac{1}{2}\nu\rangle$ that is [NiO$_6$]$^{9-}$-centers corresponding to the pseudospin projection $M$\,=\,0, whereas negative values $U$\,<\,0 stabilize the disproportionated system of spin-charge [NiO$_6$]$^{10-,8-}$-centers, corresponding to the pseudospin projection $M$\,=\,$\pm$1. 
In accordance with experimental indications on the observation of JT effect for well isolated Ni$^{3+}$-ions in LaAlO$_3$;Ni$^{3+}$\,\cite{LaAlO3_JT_ESR,LaAlO3_JT} below we choose a positive sign for the parameter $U$ in nickelates. 
However, at positive $U$, disproportionation is possible only at sufficiently large value of the screened Coulomb inter-site interaction, or nonlocal correlations, described by the effective Hamiltonian
\begin{equation}
	{\hat H}_{nloc} =  \frac{1}{2}V\sum_{i\neq j} {\hat \Sigma}_{iz}{\hat \Sigma}_{jz}\, ,
\label{nloc}	
\end{equation}
which is an analog of two-ion spin anisotropy, or Ising exchange. Nonlocal correlations are a driving force for classical disproportionation ("site-centered"\, charge order, or CO phase) with G-type  ordering of the spin-triplet electron and spinless hole centers, which, when only nearest-neighbor $nn$-interactions are taken into account, corresponds to the paramagnetic phase.

{\bf V. Transfer of effective boson  and the bosonic double exchange}.
Charge transfer effects play an important role in the formation of the spin-charge structure of nickelates.  The effective Hamiltonian of two-particle transport
\begin{equation}
  {\hat H}_{tr}^{(2)}=-\frac{1}{2}t_b\sum_{i\not= j} \left({\hat \Sigma}_{i+}^{2}{\hat \Sigma}_{j-}^{2}+ {\hat \Sigma}_{i-}^{2}{\hat \Sigma}_{j+}^{2}\right)
  \label{Hkin2}
\end{equation}
is equivalent to the transfer Hamiltonian of effective two-electron composite spin-triplet  bosons with configuration $e_g^2$;${^3}A_{1g}$ and transfer integral $t_b$.
Introducing the creation/annihilation operators ${\hat B}_{\mu}^{\dagger}$/${\hat B}_{\mu}$ for effective composite boson and selecting the spin component $\mu=0,\pm 1$, we rewrite the Hamiltonian ${\hat H}_{tr}^{(2)}$ as follows
\beq
{\hat H}_{tr}^{(2)}=-t_b\sum_{i\not= j,\mu} {\hat B}_{i\mu}^{\dagger}{\hat B}_{j\mu} \, .
\label{Hb}
\eeq
 At variance with the classical correlation contributions (\ref{loc}) and (\ref{nloc}) the quantum transfer operator ${\hat H}_{tr}^{(2)}$ does not preserve the projection of the local pseudospin $\Sigma_{iz}$, i.e., the local charge state. In other words, this operator, without changing the projection of the total pseudospin (total charge), leads to a charge density transfer with mixing of local charge states with pseudospin projections M\, =\,$\pm$1, the appearance of uncertainty of the charge state of NiO$_6$ clusters with average charge (valence) [NiO$_6$]$^{9\pm\delta}$ (Ni$^{3\pm\delta}$) and the formation of the CDq-phase of a quantum disproportionation.

Indeed, taking into account
two-particle transport in the molecular field approximation leads to the formation of local quantum superpositions
\beq
|\delta\rangle = cos\alpha\,|+1\rangle + sin\alpha\,|-1\rangle \, ,
\label{+-}
\eeq
where $\delta = \langle\Sigma_z\rangle$\,=\,$cos2\alpha$. Naturally, the quantum superpositions (\ref{+-}) at $|\delta |<\,1$ are fundamentally different from the classical states with the corresponding charge density. Thus, at $\delta$\,=\,0 we deal with an on-site superposition of  Ni$^{2+}$- and Ni$^{4+}$-centers, and not with the  Ni$^{3+}$-center. To distinguish classical and quantum states with formally the same value of $\delta$ we can use the value of the local order parameter  $\langle\Sigma^2_z\rangle$, equal to one for any superposition (\ref{+-}) and equal to zero for the Ni$^{3+}$-center, which corresponds to the pseudospin projection  M\,=\,0. 
\begin{figure}
    \centering
    \includegraphics[width=1.0\linewidth]{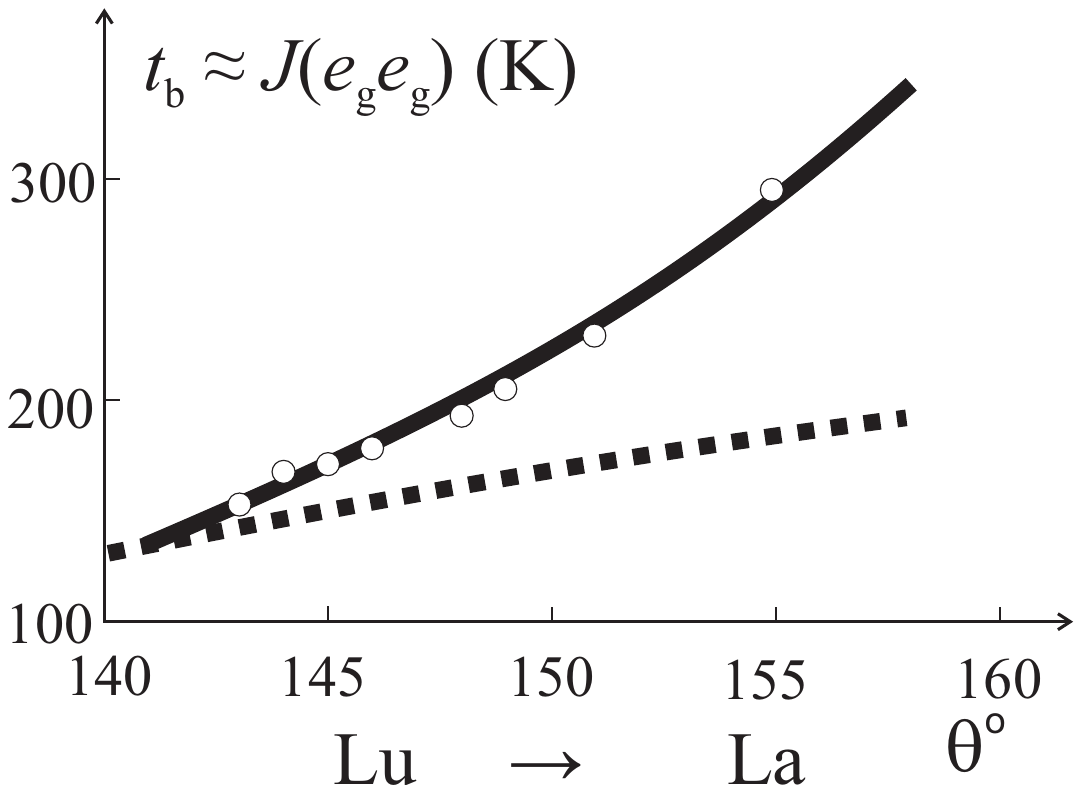}
    \caption{Dependence of the superexchange integral $J(e_ge_g)\approx t_b$ on the superexchange bonding angle, empty circles are  estimations based on the experimental data for perovskites RFeO$_3$ and RCrO$_3$\,\cite{MC_2021,JETP_2021} (line is just a guide to the eye.), dotted line shows a simplified dependence $J(e_ge_g)\propto \cos^2\theta$. Lu$\rightarrow$La  indicates the direction of growth of the superexchange bond angle in the series of nickelates.   }
    \label{fig1}
\end{figure}

The transfer of the effective local composite spin-triplet boson corresponds to the transfer of not only the charge density, but also the spin density with the conservation of the projection of the usual spin, but with the appearance of uncertainty of the on-site spin value, so that the operator ${\hat H}_{tr}^{(2)}$ is in fact also a non-traditional spin operator, or the bosonic double exchange by analogy with the traditional Zener double exchange\,\cite{Zener,Anderson_1955,DeGennes_1960}. However, this spin dependence is non-trivial. The Hamiltonian initiating the transport is spinless, which allows us to represent  ${\hat H}_{tr}^{(2)}$ in the semiclassical approximation\,\cite{Anderson_1955,Dagotto_1996} as
\beq
{\hat H}_{tr}^{(2)}=-t_b\sum_{i\not=j}  S_{ij}{\hat B}_{i}^{\dagger}{\hat B}_{j} \, ,
\label{Hb1}
\eeq
where 
$S_{ij}$ is the overlap integral of the spin functions in a common coordinate system, 
which in the simplest case can be related to the angle $\theta_{ij}$ between the spin/magnetic moments ${\bf S}_i$ and ${\bf S}_j$\,\cite{Moskvin_2025}:
\beq
S_{ij}=\cos^2\frac{\theta_{ij}}{2}\, .
\eeq
 The factor $S_{ij}$ is obviously maximal for ferromagnetic orientation of magnetic moments of neighboring sites, that is traditionally associated with the ferromagnetic nature of double exchange and attempts to introduce an effective spin Hamiltonian of the Heisenberg type. However, the transfer Hamiltonian does not allow separation of charge and spin degrees of freedom. The appearance of quantum uncertainty of the on-site spin value  with the local spin density in superpositions (\ref{+-})
\beq
\rho_s=sin^2\alpha=\frac{1\pm |\delta|}{2} 
\label{rhos}
\eeq
indicates the fundamental impossibility of associating the transfer operator with an effective spin Hamiltonian, as is often done in the theory of traditional ("single-particle") Zener double exchange\,\cite{Zener,Anderson_1955,DeGennes_1960,Dagotto_1996}.

Two-particle transfer in all JT magnets can be realized through two successive single-particle processes with the transfer of an $e_g$ electron/hole, so the composite boson transport integral $t_b$ can be estimated as follows:
\begin{equation}
t_b\approx t_{e_ge_g}^2/\Delta_{e_ge_g} \approx J_{kin}(e_ge_g) \, ,
\label{t_B}
\end{equation}
where $t_{e_ge_g}$ is the integral of the single-particle $e_g$-$e_g$ transfer, $\Delta_{e_ge_g}$ is the average transfer energy, $J_{kin}(e_ge_g)$ is the kinetic contribution to the Heisenberg superexchange integral Ni$^{2+}$-O-Ni$^{2+}$.
The dependence of the integral $J(e_ge_g)$ on the angle $\theta$ of the cation-anion-cation superexchange coupling, obtained using data for the perovskites RFeO$_3$ and RCrO$_3$ (see, e.g.\,\cite{MC_2021,JETP_2021}), is shown in Fig.\,\ref{fig1}.
Note the relatively large value of $J(e_ge_g)$, as well as  the fact that, taking into account the known data on the $\theta$  angles in nickelates\,\cite{Medarde_1997,Klein}, the exchange integral $J(e_ge_g)$, and hence the transfer integral of the effective composite boson $t_b$, will increase approximately twofold upon the transition from LuNiO$_3$ to LaNiO$_3$.
By the way, using the popular approximation $ J(e_ge_g)\propto \cos^2\theta$\,\cite{Medarde_1997,MC_2021,JETP_2021} gives in the same range of angles $\theta$ an increase in the integral $ J(e_ge_g)$ by only 20\%. Let us pay attention to the discrepancy in the data on the angles of the Ni-O-Ni superexchange bond in the papers\,\cite{Medarde_1997,Klein}, which is especially significant for "heavy"\, nickelates.

The comparison of matrix elements, or more precisely, the average values of the operators of nonlocal correlations  and boson transfer in the state $|\delta_1\delta_2\rangle$ of a pair of nearest centers described by quantum superpositions (\ref{+-})
\beq
\langle\delta_1\delta_2|{\hat H}_{nloc}|\delta_1\delta_2\rangle =V\,\cos2\alpha_1\cos2\alpha_2 \, ,
\eeq
\beq
\langle\delta_1\delta_2|{\hat H}_{tr}^{(2)}|\delta_1\delta_2\rangle =-\frac{1}{2}t_b S_{12}\,\sin2\alpha_1\sin2\alpha_2\,  
\eeq
indicates a puzzling competition of these two charge interactions.
The minimal energy (-$V$) for nonlocal correlations is realized only for CO-phase with site-centered charge order ($\alpha_1$\,=\,0, $\alpha_2$\,=\,$\frac{\pi}{2}$ or vice versa $\alpha_2$\,=\,0, $\alpha_1$\,=\,$\frac{\pi}{2}$), while the minimal transport energy (-$\frac{1}{2}t_b$) is realized for the limiting quantum superpositions with $\alpha_1$\,=\,$\alpha_2$\,=\,$\frac{\pi}{4}$ (CDq-phase with bond-centered charge order), but $S_{12}$\,=\,1, that is, for ferromagnetic spin orientation. By the way, the presence of the spin factor $S_{12}$ indicates a significant reduction in the energy of boson transfer in the spin-paramagnetic phase.

Thus, in contrast with the non-local correlations, the two-particle, or bosonic transfer is the driving force for formation of the quantum CDq-phase with the average, but quantum-mechanically uncertain values of the charge and spin for NiO$_6$-centers,  described by quantum superpositions (\ref{+-}). The charge and spin density transfer in CDq-phase  does activate  the antiferromagnetic  $nn$-superexchange Ni$^{2+}$-O$^{2-}$-Ni$^{2+}$,   forbidden in the CO phase, as well as a novel mechanism of ferromagnetic bosonic double exchange.      

The character of the charge-disproportionated state in nickelates will be determined by the CO--CDq competition due to  two competing inter-site interactions, the potential energy ${\hat H}_{nloc}$ (\ref{nloc}) and the kinetic energy of composite bosons ${\hat H}_{tr}^{(2)}$.
Indeed, in the CO phase the two-particle transfer "does not work"\, and, conversely, in the extreme quantum case $\delta$=\,0 the contribution of nonlocal correlations is switched off. 
Perhaps the most intriguing feature of the quantum transport of composite spin-triplet bosons in the CDq phase is the pespective of forming unique phase states like spin-triplet superconductivity or "supersolid"\,\cite {RMP}. In this connection, we note the recent discovery of superconducting properties in mixed-valence nickelates La$_3$Ni$_2$O$_7$\,\cite{327}.

\begin{figure}
  \includegraphics[width=1.0\linewidth]{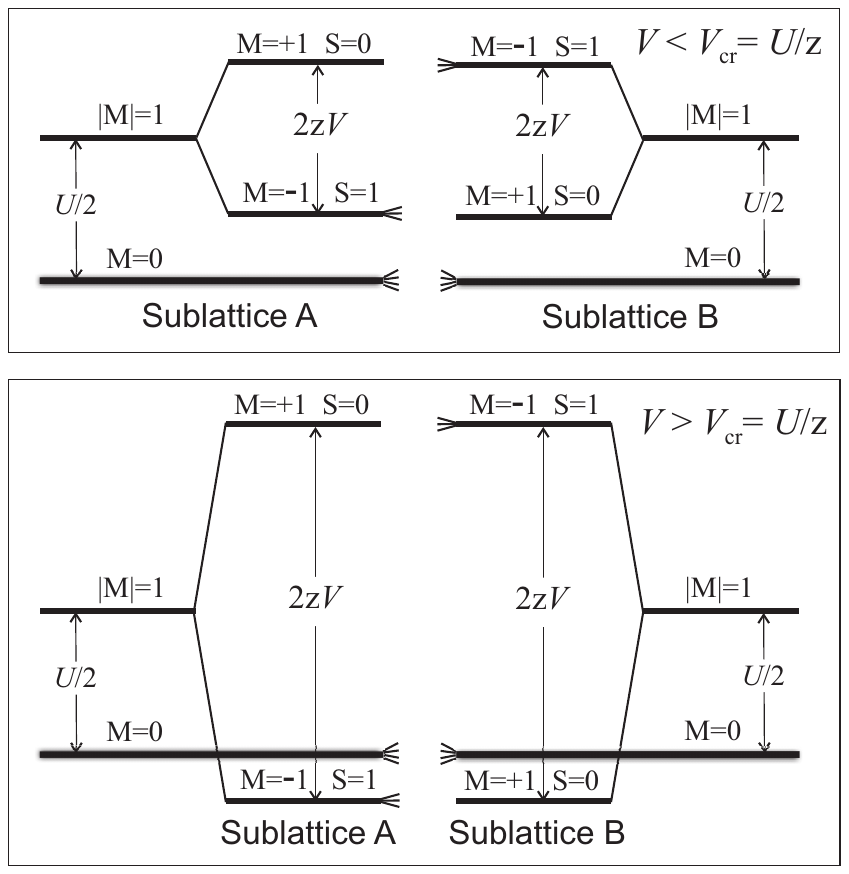}
 \caption{Scheme of the energy spectrum of octets of NiO$_6$ clusters in two sublattices of the model orthonickelate in the ground state: upper panel $V<V_{cr}$, lower panel $V>V_{cr}$.
}
\label{fig2}
\end{figure}


{\bf VI. Effective field theory.} The charge triplet model and pseudospin formalism point to a possibility of describing charge states in nickelates  using methods well known in the theory of spin magnets, first of all, the simple effective field theory (EF) to be a generalization of the mean field theory, which is a good starting point for a physically clear semi-quantitative description of strongly correlated systems\,\cite{ASM_CM_2021}.
In the effective field approximation, all local (on-site) interactions are exactly taken into account, and all inter-site interactions are taken into account within the framework of the molecular field approximation (MFA). 

Below we restrict ourselves to considering the two-sublattice ($A$ and $B$) model, introducing two charge order parameters of ferro- and antiferro-type:
\beq
\Delta n = \frac{1}{2}(\langle\langle {\hat \Sigma}_{zA}\rangle\rangle + \langle\langle {\hat \Sigma}_{zB}\rangle\rangle  )\,,\,
l=\frac{1}{2}(\langle\langle {\hat \Sigma}_{zA}\rangle\rangle  - \langle\langle {\hat \Sigma}_{zB}\rangle\rangle  ) \,,
\eeq
where $\langle\langle ...\rangle\rangle $ is a thermodynamic average.
We address pristine nickelates with half-filling, or $\Delta n$\,=\,0.

Without dwelling on the rather routine and well-known presentation of the elements of the effective field theory for the simplified $U$-$V$-$t_b$-model under consideration (see Supplemental Materials), we will move on to the presentation of the main results.

{\bf  VII. Main results and discussion}.
 Fig.\,\ref{fig2} shows the model energy spectrum of octets of Ni-centers in two sublattices of the model nickelate in atomic limit. Obviously at  $V$<$V_{cr}$\,=\,$U$/$z$, where $z$ is a coordination number (=\,6 for cubic perovskite) the ground state of Ni-centers corresponds to JT quartet (see upper panel), while at $V$>$V_{cr}$  the ground state of Ni-centers corresponds to electron(hole) center in A(B) sublattice (see bottom panel). The spectrum in Fig.\,\ref{fig2} allows us to find the energies of  single-  and  two-particle  optical charge transfer  transitions in the CO phase of nickelates ($\Delta_1$\,=\,z$V$\,-\,$U$ and $\Delta_2$\,=\,4z$V$, respectively) and estimate the parameters of local and nonlocal correlation. Unfortunately, the available optical data for nickelates\,\cite{optics} are extremely scarce.

\begin{figure}
\centering
\includegraphics[width=1.0\linewidth]{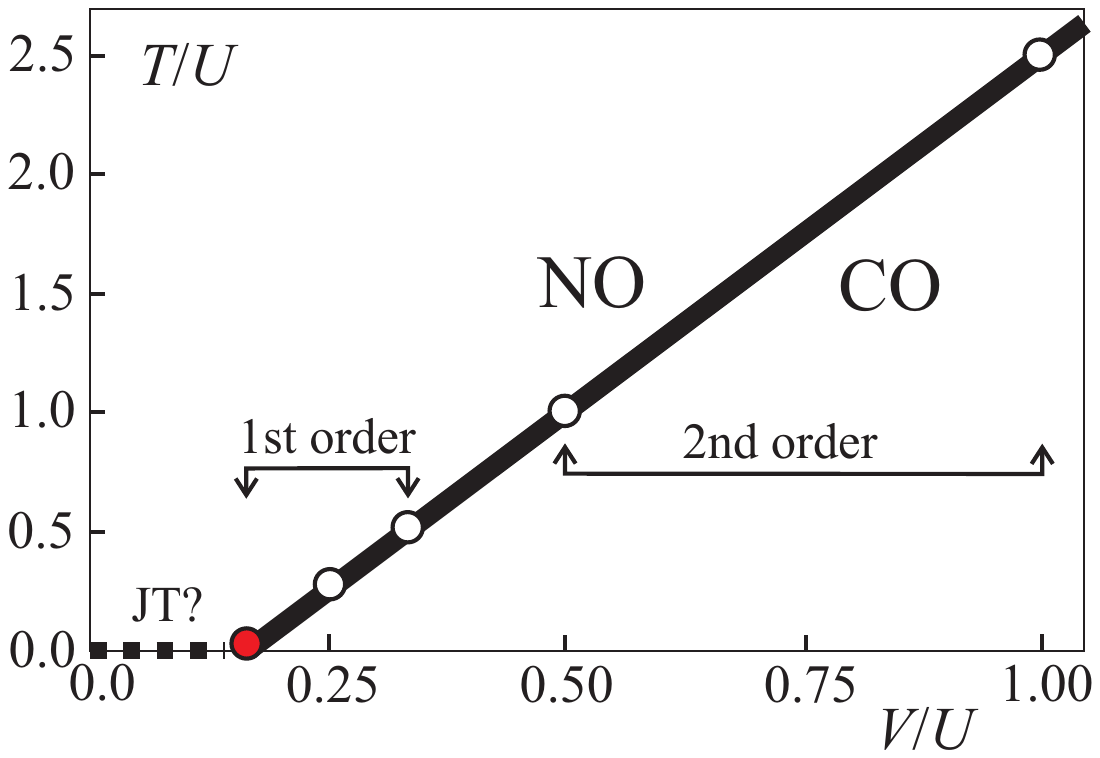}
\caption{Dependence of the CO--NO transition temperature on the nonlocal correlation parameter in the atomic limit of the model ($T-V$-phase diagram).  Line is
just a guide to the eye.}
\label{fig3}
\end{figure}

Using the expression for the on-site free energy, we calculated the dependence $f(l)$ on the order parameter $l$ for different temperatures and different values of the parameter $V$ in units of $U$\,>\,0, given $t_b$\,=\,0 (see Exp.\,(S7) and Fig.\,S1 in Supplemental Material\,\cite{SM}) and constructed the $T$-$V$ phase diagram of model nickelate in the atomic limit ($U$-$V$-model), shown in Fig.\,\ref{fig3}. Here the line $T(V)$ separates the low-temperature insulating CO-phase with classical charge disproportionation and the high-temperature non-ordered NO-phase, which we associate with the bad metal phase.

The CO-NO transition is realized only at $V$>$V_{cr}$\,=\,$\frac{1}{z}U$\,=\,$\frac{1}{6}U$, so that at $V$\,=\,$V_{cr}$ and $T$\,=\,0 the free energies of the CO and NO phases coincide, the transition temperature turns into zero, $T_{\mbox{{\footnotesize CO}}}$\,=\,0. At $V$<$V_{cr}$ the Jahn-Teller cooperative ordering may be realized. At $V$>$V_{cr}$ the CO--NO transition temperature increases almost linearly with increasing  parameter $V$. 
Analysis of the curves  $f(l)$ (see Fig.\,S1 in Supplemental Material\,\cite{SM}) shows that this dependence  at $V_{cr}$<$V$$\leq$\,$\frac{1}{3}U$ is typical for the first-order phase transitions, while at $V$$\geq$\,$\frac{1}{2}U$ it is typical for second-order phase transitions. In the intermediate region  $\frac{1}{3}U$<$V$<\,$\frac{1}{2}U$ the features typical for both second- and first-order phase transitions are observed. The high-temperature NO-phase is stable over the entire temperature range down to the lowest temperatures. At the same time, the low-temperature insulating CO-phase becomes stable significantly above the critical temperature $T_{\mbox{{\footnotesize CO}}}$, that is the maximal temperature at which the free energies of the NO and CO phases coincide. Thus, phase separation can be considered a typical property of nickelates.


\begin{figure*}
\centering
\includegraphics[width=1.0\linewidth]{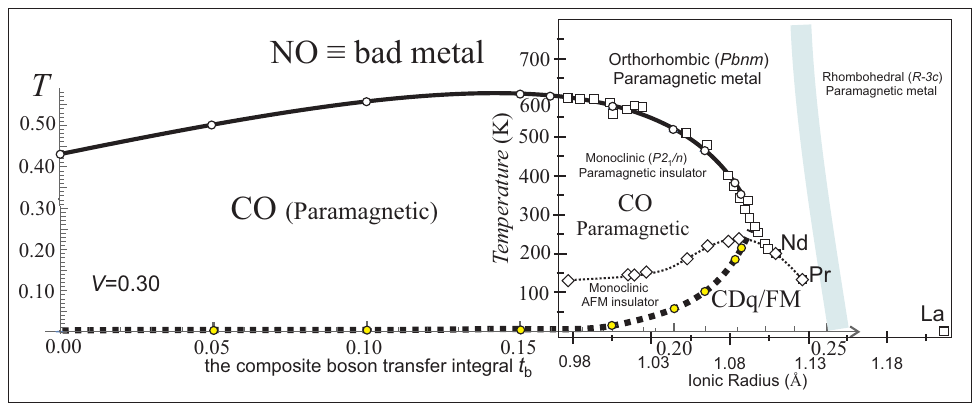}
\caption{Model $T$-$t_b$-phase diagram for nickelates. Dependence of critical temperatures of CDq-CO (filled circles and bold dotted line) and CO-NO (empty circles and solid bold line) transitions on the value of the composite boson transfer integral $t_b$ and comparison with the experimental data on $T_{\mbox{{\footnotesize MIT}}}$ (squares) and $T_N$ (rhombuses and dotted line) and indications of crystal, electronic and magnetic structure for different orthonickelates RNiO$_3$\,\cite{Gavrilyuk}.  Lines are
just a guide to the eye, $T$, $V$, and $t_b$ are in units of $U$.}
\label{fig4}
\end{figure*}

The inclusion of the quantum effect of two-particle boson transfer ${\hat H}_{tr}^{(2)}$ within the $U$-$V$-$t_b$-model leads to a fundamental  rearrangement of both the ground state, phase transitions, and the temperature phase diagram of nickelates\,\cite{SM}. At sufficiently small values of the transfer integral $t_b$ compared to the parameter $V$, the temperature range of existence of the CDq phase turns out to be small, so that a small increase in temperature leads to a CDq--CO phase transition  to the classical disproportionation phase with a subsequent CO--NO transition  to the bad-metallic NO phase. However, the competition of the potential and kinetic energy of effective composite bosons, that is, nonlocal correlations and pair transfer, leads to an unexpected effect. 
With increasing value of the boson transfer integral, the temperature $T_{\mbox{{\footnotesize CDq}}}$ of the CDq--CO transition first increases slowly, being close to zero, and then at some critical value $t_b^s$ it begins to grow sharply, while the temperature $T_{\mbox{{\footnotesize CO}}}$ of the CO-NO transition to the bad-metallic phase, i.e. the insulator-metal transition temperature, first increases slowly, reaches a maximum, then at $t_b\approx t_b^s$ it begins to fall sharply, and at some critical value  $t_b^*$ the temperatures $T_{\mbox{{\footnotesize CO}}}$ and $T_{\mbox{{\footnotesize CDq}}}$ become equal, and at $t_b$\,>\,$t_b^*$ only the CDq phase survives. 
The parameter |$\delta$|, which determines local quantum superpositions (\ref{+-}), increases slightly with increasing $t_b$ over a wide interval, remaining close to 1, and then at $t_b\sim t_b^s$ it starts to fall sharply, turning to zero at $t_b\sim t_b^*$.

All these features are well illustrated in Fig.\,\ref{fig4}, being our main result, where we present a $T$-$t_b$ phase diagram given $V$\,=\,0.30, that corresponds to the range of weakly first-order CO--NO transitions. 
Under the simplest assumption of a linear relationship between the integral $t_b$ and the ionic radius for the R-ions in the range of its real values for orthonickelates, the dependence $T(t_b)$ perfectly reproduces the R-dependence of the paramagnetic insulator-bad metal transition temperatures for orthonickelates RNiO$_3$ with
$T_{\mbox{{\footnotesize MIT}}}$\,$\neq$\,$T_{\mbox{{\footnotesize N}}}$ (R= Lu, ..., Sm)\,\cite{Gavrilyuk} 
 given $U$$\approx$\,1000\,$K$, that immediately yields $V$$\approx$\,300\,$K$. In this case, the value of the boson transfer integral changes from $t_b$\,$\approx$\,165\,$K$, corresponding to LuNiO$_3$, to $t_b$\,=\,$t_b^*$\,$\approx$\,225\,$K$, approximately corresponding to SmNiO$_3$, which, taking into account the corresponding  superexchange angles\,\cite{Medarde_1997}, is in good qualitative and quantitative agreement with the theoretical predictions shown in Fig.\, \ref{fig1}.
At $t_b$\,>\,$t_b^*$, our model, in agreement with experiment, indicates a fundamental change in the character of the MIT  from the NO $\rightarrow$ classical paramagnetic CO phase to the NO $\rightarrow$ quantum magnetic CDq phase. However, the observed complex non-collinear antiferromagnetic structure ${\bf k}$\,=\,(1/2,0,1/2)\,\cite{Medarde_1997,Catalano_2018} of this phase and the unusual dependence of $T_{\mbox{{\footnotesize MIT}}}$(R)\,=\,$T_{\mbox{{\footnotesize N}}}$(R)  for R\,=\,Nd, Pr and close compositions indicate the limitations of the minimal purely charge model and the need to take into account the very unusual competition of ferromagnetic bosonic double exchange and antiferromagnetic superexchange\,\cite{SM}, that is, the transition to the $U$-$V$-$t_b$-$J$-model.
A more realistic model should also take into account a noticeable decrease in the value of the nonlocal correlation parameter $V$ in the LuNiO$_3$ -- LaNiO$_3$ series.  




{\bf VIII.\, Conclusion.} We have presented a novel purely electronic scenario  of  insulator-to-bad metal transition in the nickelates RNiO$_3$ based on a minimal $U$-$V$-$t_b$-model which takes into account the only charge degree of freedom within a pseudospin formalism and  effective field theory that provides a physically clear and mathematically obvious description typical for traditional spin-magnetic systems. 


The minimal model differs fundamentally from the traditional Hubbard model, starting from the physical model itself, the local (on-site) Hilbert space, the effective Hamiltonian with the CD ground state, and the conclusion about the fundamental role of nonlocal correlations  and two-particle transport  for the description of the charge degree of freedom and features of the temperature phase diagram, as well as the $T_{\mbox{{\footnotesize MIT}}}$(R) dependence. Let remind that traditional Hubbard model considers  MIT  as a result of the competition of the band width $W$ defined by the single-particle transfer integral $t$ (integrals in the two-band Hubbard models) and the value of local correlations $U$. 
Time-consuming DFT-based techniques   can provide a plausible semi-quantitative description of some features of the MIT in nickelates, however, its precise quantitative accuracy is unknown\,\cite{Millis_2022}  both because of doubts about the adequacy of the DFT description of bare uncorrelated orbitals\,\cite{Millis_2022} and because of the unsolved problem of adequately accounting for local and nonlocal correlations. 

Despite the over-simplicity of the model it is believed to reproduce main features of the $T$-$R$-phase diagram for RNiO$_3$ under physically reasonable estimations for the parameters of the $U$-$V$-$t_b$-model.
High-temperature bad-metallic phase is associated with the non-ordered mixed-valence NO phase of the charge triplet system, while   the insulating disproportionated CD-phase is composed of the low-temperature quantum magnetic CDq-phase and high-temperature classical paramagnetic CO-phase separated by the CDq-CO phase transition (R\,=\,Lu,...,Sm) or only magnetic CDq-phase (R\,=\,Nd, Pr). 

For a detailed description of the quantum CDq phase it is necessary to go beyond the purely charge model taking into account the unusual competition of ferromagnetic bosonic double exchange and antiferromagnetic superexchange Ni$^{2+}$-O$^{2-}$-Ni$^{2+}$.


Preliminary analysis shows that the pseudospin formalism makes it possible to effectively take into account the electron-lattice interaction, first of all the most important contribution of the so-called "breathing"\, mode of local distortions of NiO$_6$ centers and bipolaronic effects, as well as other manifestations of the important role of electron-lattice coupling experimentally observed in nickelates\,\cite{Catalano_2018}. 
Analysis of the electron-lattice interplay, magnetic structure and  superconducting perspectives of the quantum CDq phase will be a matter of the near future work.

The author is grateful to Yu. Panov for valuable discussions.

{\bf Funding}.
This research was supported  by the Ministry of Science and Higher Education of the Russian Federation within the framework of the Project FEUZ-2023-0017 (main text) and state assignment  for the M.N. Mikheev lnstitute of Metal Physics of Ural Branch of Russian Academy of Sciences (Supplemental Material).

{\bf Conflict of interest}. 
The author declares that he has no conflict of interest.

\end{document}